\documentclass{article}

\usepackage[backref=page]{hyperref}

\usepackage[latin3]{inputenc}
\usepackage{amsmath}
\usepackage{amsfonts}
\usepackage{amssymb}
\usepackage{amsthm}
\usepackage{mathrsfs}
\usepackage{tikz}

\usepackage[authoryear]{natbib}

\newcommand\B{\mathbb B}
\newcommand\R{\mathbb R}
\newcommand\J{\mathscr J}
\newcommand\K{\mathscr K}
\renewcommand\epsilon\varepsilon
\newcommand\callcc{\text{c}\!\text{c}}
\newcommand\bind{>\!\!>\!\!=}
\newcommand\argmax{\operatorname{arg\,max}}
\newcommand\ignore[1]{}

\title{The selection monad as a CPS translation}
\author{Jules Hedges}
\date\today

\begin{document}

\pagestyle{plain}

\pagenumbering{roman}

\maketitle
\begin{abstract}\noindent
A computation in the continuation monad returns a final result given a continuation, ie. it is a function with type $(X \to R) \to R$. If we instead return the intermediate result at $X$ then our computation is called a selection function. Selection functions appear in diverse areas of mathematics and computer science (especially game theory, proof theory and topology) but the existing literature does not heavily emphasise the fact that the selection monad is a CPS translation. In particular it has so far gone unnoticed that the selection monad has a call/cc-like operator with interesting similarities and differences to the usual call/cc, which we explore experimentally using Haskell.

Selection functions can be used whenever we find the intermediate result more interesting than the final result. For example a SAT solver computes an assignment to a boolean function, and then its continuation decides whether it is a satisfying assignment, and we find the assignment itself more interesting than the fact that it is or is not satisfying. In game theory we find the move chosen by a player more interesting than the outcome that results from that move. The author and collaborators are developing a theory of games in which selection functions are viewed as generalised notions of rationality, used to model players. By realising that strategic contexts in game theory are examples of continuations we can see that classical game theory narrowly misses being in CPS, and that a small change of viewpoint yields a theory of games that is better behaved, and especially more compositional.
\end{abstract}

\clearpage

\ 

\vfill 

\tableofcontents

\vfill 

\ 

\clearpage

\pagenumbering{arabic}


\section{Introduction}

A selection function is a type-2 function $\varepsilon : \J X$ where
\[ \J X = (X \to R) \to X \]
We can view the input $k : X \to R$ to such a function in several ways: as a generalised predicate, as the context of a decision, or as a continuation. These are emphasised respectively in topology \citep{escardo08}, game theory \citep{escardo11} and proof theory \citep{escardo12b}. The earliest selection functions considered were computable instances of Hilbert's $\varepsilon$-operator $(X \to \B) \to X$, which witness a computational form of topological compactness. An operation $\otimes : \J X \times \J Y \to \J (X \times Y)$, which is the (left-leaning) monoidal product of the strong monad $\J$, witnesses the fact that the product of two compact spaces is compact. Remarkably this extends to countable products, leading to the derivation of `seemingly impossible functional programs' \citep{escardo07} that search set-theoretically infinite but topologically compact types such as $\mathbb N \to \B$ (the Cantor space) in a finite amount of time.

It was noticed by Paulo Oliva that the product of selection functions is equivalent to Spector's bar recursion \citep{escardo14}, a notoriously obscure computational feature used to realise the axiom of countable choice via a double negative translation and G\"odel's Dialectica interpretation. Bar recursion is important in the proof mining programme because it can be used to interpret proofs of classical analysis (including differential equations and ergodic theory) \citep{kohlenbach08}, but the computer programs arising from such proofs via bar recursion, while provably correct, are not well suited to human understanding.

The next step made by Escard\'o and Oliva was the connection with game theory, by realising that economic rationality is modelled by the selection function
\[ \argmax : (X \to \R) \to X \]
which finds a point maximising a real-valued function (say, on a finite set $X$). By generalising selection functions by replacing the booleans with $\R$ or an arbitrary type $R$, the product of selection functions is seen to correspond to an well-known and intuitive algorithm in game theory known as backward induction \citep{escardo11}. Since proof interpretations using bar recursion can be rewritten using the product of selection functions, we therefore obtain a computational interpretation of proofs in classical analysis that is amenable to human understanding via game theory \citep{oliva12b}.

The relationship to the continuation monad is immediately obvious: the Hilbert $\epsilon$-operator is a refined form of the quantifier
\[ \exists : (X \to \B) \to \B \]
and the operator $\argmax$ is a refined form of
\[ \max : (X \to \R) \to \R \]
These functions are both computations in the continuation monad
\[ \K X = (X \to R) \to R \]
Escard\'o and Oliva call any function with a type of this form a \emph{quantifier}. Moreover the relationships
\[ \exists p = p (\epsilon p) \]
and
\[ \max p = p (\argmax p) \]
define a monad morphism from the selection to the continuation monad. Thus in proof theory the J-translation (or Peirce translation) \citep{escardo12b} can be seen as a refined form of the usual (generalised) double negative translation, and it would be tempting to view the selection monad as a refinement of the continuation monad. However things are not so clear, because call/cc of the selection monad has a more specific type, and behaves differently.

In section \ref{monads} we will explain the (quite complex) bind operation of the selection monad using intuition from CPS, which previously has only been derived in a purely formal way. In sections \ref{callcc1} and \ref{callcc2} we will discuss the call/cc-like operator that exists for the selection monad, and see its behaviour experimentally. Finally in section \ref{games1} we will informally discuss ongoing work by the author and others on applications of selection functions and CPS in game theory. 

\section{Two monads for CPS}\label{monads}

We begin by recalling the usual intuition for the continuation monad, which can be used to implement (delimited) continuations in a pure language such as Haskell. We view all computation as being done relative to a \emph{continuation}, which takes the return value and chains it to the future of the computation. Thus our computation has the shape\footnote{The diagrams in this paper are not intended to be formal, but rather as a possible aid to intuition, which not every reader will find helpful. `Plain' arrows are intended to live in the Kleisli category of the base monad $M$, while arrows with quote marks around the name live in the Kleisli category of either $\K^M_R$ or $\J^M_R$.}
\begin{center}\begin{tikzpicture}[node distance=3cm, auto]
\node (A) {$\cdots$};
\node (X) [right of=A] {$X$};
\node (R) [right of=X] {$R$};
\draw [->] (A) to node {} (X);
\draw [->] (X) to node {$k$} (R);
\end{tikzpicture}\end{center}
We will call $X$ the type of \emph{intermediate values}, and $R$ the type of \emph{final values}. The key idea of the continuation monad is that we always work relative to an \emph{unknown} continuation $k$, although we may fix the type $R$. (The ability to fix $R$ may have begun as a quirk due to Hindley-Milner typing, see \citep{kiselyov}, but it is vital to many applications of selection functions.) We allow our functions to have side-effects, modelled by functions $X \to MR$ for a suitable monad $M$. A computation in the continuation monad is a function with type
\[ \K^M_R = (X \to MR) \to MR \]
which computes a final result given a continuation.

To embed a pure value $x : X$ we simply view it as the intermediate value, and apply the continuation to it immediately:
\[ \eta_{\K^M_R} x = \lambda k^{X \to MR} . kx \]
For the bind, suppose we have a computation $\varphi : \K^M_R X$ and a family of computations $F : X \to \K^M_R Y$. Equivalently, $F$ is a computation $X \to Y$, which we can run by supplying it with a continuation from $Y$. We form a computation that has the shape
\begin{center}\begin{tikzpicture}[node distance=3cm, auto]
\node (A) {$\cdots$};
\node (X) [right of=A] {$X$};
\node (Y) [right of=X] {$Y$};
\node (R) [right of=Y] {$R$};
\draw [->] (A) to node {} (X);
\draw [->] (X) to node {``$F$''} (Y);
\draw [->] (Y) to node {$k$} (R);
\end{tikzpicture}\end{center}
The key to understanding the bind of both the continuation and selection monads is that we have two views of a computation like this: we can either view the intermediate result as being at $X$ or $Y$. The `external' view of $\varphi \bind_{\K^M_R} F : \K^M_R Y$ is that it is a computation with the intermediate result being at $Y$. Suppose we are running this computation, so we have a continuation $k : Y \to MR$. To find the final result we move to the other view, and run the computation $\varphi$ by building it a longer continuation $k' : X \to MR$. Given an intermediate value $x : X$ we can now find the final result $k'x$ because we have the continuation $k$ from $Y$: it is $k'x = Fxk$. Therefore the bind operator for the continuation monad is given by
\[ \varphi \bind_{\K^M_R} F = \lambda k^{Y \to MR} . \varphi \lambda x^X . Fxk \]

This intuition for the continuation monad transfers directly to the selection monad. Whereas a computation in the continuation monad computes the final result given a continuation, a computation in the selection monad computes the intermediate result instead. Thus the selection monad is
\[ \J^M_R X = (X \to MR) \to MX \]

First, suppose we want to embed a pure value $x : X$ as the intermediate result. Given a continuation $k : X \to MR$, we ignore the continuation and simply return $x$, embedded as a computation in $M$:
\[ \eta_{\J^M_R} x = \lambda k^{X \to MR} . \eta_M x \]

For the bind operator of the selection monad we must again move between the two views of the computation
\begin{center}\begin{tikzpicture}[node distance=3cm, auto]
\node (A) {$\cdots$};
\node (X) [right of=A] {$X$};
\node (Y) [right of=X] {$Y$};
\node (R) [right of=Y] {$R$};
\draw [->] (A) to node {} (X);
\draw [->] (X) to node {``$F$''} (Y);
\draw [->] (Y) to node {$k$} (R);
\end{tikzpicture}\end{center}
Suppose we have a computation $\epsilon : \J^M_R X$ and a family of computations $F : X \to \J^M_R Y$. To run $\epsilon \bind_{\J^M_R} F : \J^M_R Y$ we are given a continuation $k : Y \to MR$, and we must return an intermediate result at $Y$. Our first task is to find a way to turn an intermediate result at $X$ into one at $Y$. Suppose we have an intermediate value $x : X$, so we have a computation $Fx : \J^M_R Y$ with intermediate value at $Y$. We can now run this computation with our continuation $k$, which yields an intermediate value at $Y$. Thus we have a function $f : X \to MY$ given by
\[ fx = Fxk \]
Now we can extend our continuation $k$ with $f$ to give a continuation $k'$ from $X$, namely
\[ k'x = fx \bind_M k = Fxk \bind_M k \]
By running the computation $\epsilon$ with this continuation, we obtain an intermediate result at $X$. Finally, we must apply the function $f$ again to obtain an intermediate result at $Y$, which is what we need. Written out explicitly, the operator is
\[ \epsilon \bind_{\J^M_R} F = \lambda k^{Y \to MR} . (\epsilon \lambda x^X . Fxk \bind_M k) \bind_M \lambda x^X . Fxk \]

Obviously, the sheer complexity of this formula makes reasoning about programs in the selection monad very difficult. In practice, Haskell is a very useful tool for qualitatively understanding the behaviour of such programs.

No published proof of the monad laws for the selection monad exists. The original proof (in the absence of other side-effects) was generated by an equational reasoning tool written by Martin Escard\'o specifically for this purpose, resulting in several pages of formal manipulations. For an arbitrary side-effects the unit laws were checked by the author with several pages of equational reasoning, including use of functional extensionality ($\eta$-expansion). The associative law seems to be impractical to check by hand, but the proof is found by Coq's tactic for intuitionistic logic. However previously the unit and bind operators were considered purely formal objects (derived simply by proving them as theorems of intuitionistic logic), and the intuitions developed in this section suggest for the first time that it might be possible to find a human-readable proof, by reducing to the monad laws for the continuation monad.

\section{Call/cc for the selection monad}\label{callcc1}

For building computations in the continuation monad we can use the call-with-current-continuation operator. This has type
\[ \callcc : ((X \to \K^M_R Y) \to \K^M_R X) \to \K^M_R X \]
The input $\Phi$ to $\callcc$ is called a \emph{continuation handler}, which is a computation that has access to the \emph{current continuation} $\Bbbk : X \to \K^M_R Y$ (we will use the letter $k$ to refer to an actual continuation, and $\Bbbk$ to refer to a continuation reified as a computation in the continuation or selection monad). The current continuation is just a computation $X \to Y$ in the continuation monad, and the purpose of $\callcc \Phi$ is to call $\Phi \Bbbk$ where $\Bbbk$ is bound to the continuation of $\callcc \Phi$. The implementation of $\callcc$ is given by
\[ \callcc \Phi = \lambda k_1^{X \to MR} . \Phi (\lambda x^X, k_2^{Y \to MR} . k_1 x) k_1 \]
Given our description it is quite easy to read this formula. The parameter given to $\Phi$ should be the continuation $k_1$ reified as a function $X \to K^M_R Y$. We have exactly that: given an input, we make the computation which ignores its own continuation and uses $k_1$ instead.

Put another way, we want to build a computation $\Bbbk$ to which we can apply $\Phi$. Consider the diagram
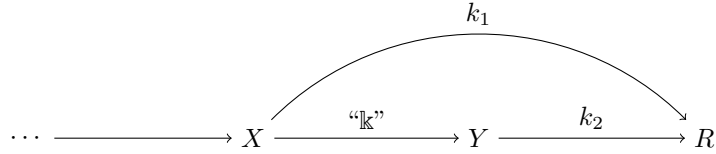
\begin{center}\begin{tikzpicture}[node distance=3cm, auto]
\node (A) {$\cdots$};
\node (X) [right of=A] {$X$};
\node (Y) [right of=X] {$Y$};
\node (R) [right of=Y] {$R$};
\draw [->] (A) to node {} (X);
\draw [->] (X) to node {``$\Bbbk$''} (Y);
\draw [->] (Y) to node {$k_2$} (R);
\draw [->] (X) edge [bend left=45] node [above] {$k_1$} (R);
\end{tikzpicture}\end{center}
In this diagram, $X$ is the intermediate type at the point in the handler at which the current continuation is invoked, and $Y$ is the intermediate result when the handler ends. We do not have a pure function $X \to Y$, but we can build a CPS computation $\Bbbk : X \to \K^M_R Y$ by short-cutting with the continuation $k_1$, that is,
\[ \Bbbk x = \lambda k_2^{Y \to MR} . k_1 x \]

However, the principal type of the term $\callcc$ is the far more general
\[ ((A \to B \to C) \to (A \to C) \to D) \to (A \to C) \to D \]
We obtain the specific type $((X \to \K^M_R Y) \to \K^M_R X) \to \K^M_R X$ by setting $A = X$, $B = Y \to MR$ and $C = D = MR$. In order that our computations are in the selection rather than the continuation monad, we instead set $C = MR$ and $D = MX$. This implies that $Y = R$, and so $B = R \to MR$. Thus $\callcc$ has the type
\[ ((X \to \J^M_R R) \to \J^M_R X) \to \J^M_R X \]
This is the call-with-current-continuation operator for the selection monad. It is equal (as an untyped $\lambda$-term) to the ordinary call/cc, and the difference in behaviour comes from the different bind operator with which it is composed.

If we draw a similar diagram we get
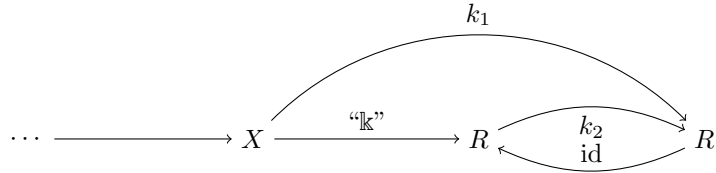
\begin{center}\begin{tikzpicture}[node distance=3cm, auto]
\node (A) {$\cdots$};
\node (X) [right of=A] {$X$};
\node (Y) [right of=X] {$R$};
\node (R) [right of=Y] {$R$};
\draw [->] (A) to node {} (X);
\draw [->] (X) to node {``$\Bbbk$''} (Y);
\draw [->] (Y) edge [bend left=25] node [below] {$k_2$} (R);
\draw [->] (R) edge [bend left=25] node [above] {id} (Y);
\draw [->] (X) edge [bend left=45] node [above] {$k_1$} (R);
\end{tikzpicture}\end{center}
We have the continuation $k_1$ which gives us a final result given an intermediate result at $X$, which is also the point at which the continuation is invoked in the handler. The computation $\Bbbk$, given its continuation $k_2$, should return the intermediate result. However instead we simply return $k_1 x$, which is really the final result, which we arranged to have the same type. That is, we are implicitly assuming that $k_2$ is always the trivial continuation, even though it might not be. We could change the type to
\[ ((X \to \K^M_R R) \to \J^M_R X) \to \J^M_R X \]
to emphasise that $\Bbbk$ returns a final rather than an intermediate result, but that only moves our dishonesty elsewhere because to actually invoke the current continuation in the handler we would need an `invoke-continuation' function with type $\K^M_R R \to \J^M_R R$, which treats its final result as intermediate.

In the next section we will see experimental results about the operational behaviour of this function, including when the whole computation is run with a nontrivial continuation.

\section{Programming with selection call/cc}\label{callcc2}

This section uses the author's implementation of the selection monad transformer from \citep{hedges14}, and assumes some familiarity with monad transformers and the syntax of Haskell. As an example we will take the simple CPS Haskell program
\begin{verbatim}
trace :: (MonadIO m) => String -> m ()
trace = liftIO . putStrLn

foo :: ContT r IO Int
foo = do trace "In foo"
         n <- callCC $ \k -> do trace "In handler"
                                m <- k 0
                                trace "Still in handler"
                                return (m + 1)
         trace "In continuation"
         return (n + 1)
\end{verbatim}
When run interactively we obtain
\begin{verbatim}
ghci> runContT foo return
In foo
In handler
In continuation
1
\end{verbatim}

This program demonstrates an important fact about call/cc: once the continuation is run, control is never returned to the handler. Thus while the continuation logically returns a value and we can bind it to a variable, any code after the continuation is called is unreachable. This is in contrast to the selection monad, which we will see next.

The same program written in the selection monad is
\begin{verbatim}
bar :: SelT Int IO Int
bar = do trace "In bar"
         n <- callCC' $ \k -> do trace "In handler"
                                 m <- k 0
                                 trace "Still in handler"
                                 return (m + 1)
         trace "In continuation"
         return (n + 1)
\end{verbatim}
By running this program we can see the execution trace:
\begin{verbatim}
ghci> runSelT bar return
In bar
In handler
In continuation
Still in handler
In continuation
3
\end{verbatim}
As can be seen, there is a coroutine-like dialogue between the handler and the continuation, with data flowing back and forth.

Like in the continuation monad, the continuation can be called from arbitrarily far inside a call stack. In particular we can write a product of selection functions that has access to its current continuation. For example, we can write a SAT solver that calls the current continuation with a dummy input once per iteration:
\begin{verbatim}
sat :: Int -> SelT Bool IO [Bool]
sat n = do bs <- callCC' $ \k -> sequence $ replicate n $
             do b <- SelT ($ True)
                liftIO $ putStr $ "b = " ++ show b ++ ", "
                k []
                return b
           trace $ "Continuation called with " ++ show bs
           return bs
\end{verbatim}
This program is based on the verbose SAT solver in \citep{hedges14}, and is designed to exhibit as many unexplained patterns as possible. Given a particular formula like
\begin{verbatim}
f :: [Bool] -> IO Bool
f bs = return $ bs!!0 && not(bs!!1) && bs!!2
\end{verbatim}
we can run it by
\begin{verbatim}
ghci> runSelT (sat 3) f
b = True, Continuation called with []
b = True, Continuation called with []
b = True, Continuation called with []
Continuation called with [True,True,True]
b = False, Continuation called with []
Continuation called with [True,True,False]
b = False, Continuation called with []
b = True, Continuation called with []
Continuation called with [True,False,True]
b = True, Continuation called with []
Continuation called with [True,False,True]
b = True, Continuation called with []
b = True, Continuation called with []
b = True, Continuation called with []
Continuation called with [True,True,True]
b = False, Continuation called with []
Continuation called with [True,True,False]
b = False, Continuation called with []
b = True, Continuation called with []
Continuation called with [True,False,True]
b = True, Continuation called with []
Continuation called with [True,False,True]
[True,False,True]
\end{verbatim}
(Note that each time the current continuation is called with the empty list it also calls the formula $f$, but the expressions which index the empty list, which would result in a runtime exception, are not evaluated due to the lazy semantics of Haskell.) It is already an open question how to explain patterns such as $TTT, TTF, TFT, TFT, TTT, TTF, TFT, TFT$ resulting from the operational behaviour of the product of selection functions. The positions of the empty list in this trace only seem to deepen the mystery.

\section{Strategic contexts are continuations}\label{games1}

The author and collaborators are developing a theory of games based on selection functions, which will be informally described in this section. In a computable game theory, the computations done by players are naturally CPS computations, in the sense that the computation divides into two parts: first the player computes a move, and then the rules of the game (and the other players) use the move to compute the outcome. However the player's computation of a move is done with knowledge of how that move will be used to compute an outcome, which makes it a CPS computation. In classical game theory a player may have a finite set of choices $X$, and each choice determines a real number $kx : \R$ called a \emph{utility}. A fundamental assumption of classical game theory is that rational players act so as to maximise their utility. We can view the continuation of the player's decision as $k : X \to \R$, and the computation done by the player is $\argmax : \J_\R X$. One reason that this is interesting is that the product of selection functions computes Nash equilibria of sequential games of perfect information, by implementing an algorithm known as backward induction \citep{escardo12}. Thus what was apparently a fact of applied mathematics, concerning the behaviour of interacting rational agents, in fact arises very naturally in a theoretical setting.

An immediate application of these ideas is that we can easily generalise some concepts in classical game theory to arbitrary selection functions, which allows us to work with players who are not classically rational without having to build a new game theory from scratch. This is applied to `context-dependent choice' in voting games in \citep{hedges14c}, where it is shown that the fixpoint operator $\operatorname{fix} : \J^\mathcal P_X X$ models a `Keynes agent' who would like to vote for the winner of an election, or a so-called Keynes beauty contest. Similarly the operator that selects non-fixpoints models a `punk' who would like to vote for anybody but the winner, which corresponds to a one-shot minority game \citep{kets11}. In general, a `context-dependent agent' may have preferences not just over outcomes, but over the ways in which those outcomes are achieved. This formally gives no new expressive power because the type of outcomes could be extended to include all relevant information, but selection games are more convenient, and in particular seem to be more scalable. To give a silly example, a million dollars earned is not equivalent to a million dollars stolen. To model such `social concerns' classically, an explicit conversion rate between morality and dollars must be given globally and encoded into the outcome structure of the game. With selection functions this can be done instead on a per-agent basis.

We conclude by giving several research directions currently being explored by the author and several collaborators. In each of these, the slogan `strategic contexts are continuations' is an important part of the author's intuitive understanding.

\begin{itemize}
\item Most importantly, the author has recently developed a graphical calculus for game theory by extending string diagrams for monoidal categories \citep{selinger11}. The semantics of these string diagrams uses CPS very heavily: a diagram generally represents some group of interacting agents, which is equivalent to a `generalised agent' with preferences over both strategies and continuations from computations of strategies. The categorical composition and tensor product (which are primitive forms of sequential and parallel composition, respectively) are both defined using (delimited) continuation transformers, and cannot be written in a direct style.

\item The way in which this graphical calculus differs from the ones used in quantum theory is the way in which backward-causality is treated. Game theory contains a quite restricted form of backward-causality due to agents reasoning about future events. An agent is graphically connected to a relevant future value by a feedback-like operation. Intriguingly there is an extremely close analogy with shift/reset operators \citep{filinski90} here: a decision made by an agent is like `shift' (it captures a continuation), and the point in the future at which a value is designated as the outcome for an agent is like `reset' (it delimits a continuation). In the graphical calculus these also appear strongly analogous to the unit and counit of a compact structure in category theory, which correspond to pair-production and pair-annihilation in quantum mechanics.

\item The first two points relate to defining games and solution concepts, but not to computing solutions in general. In this direction, the author has used various monad transformer stacks with the selection monad at the top. For sequential games of perfect information this can be done in generality using the product of selection functions, but to extend to simultaneous and other games must be done on a per-effect basis. The most progress has been made for nondeterministic games, in which agents can make nondeterministic choices between several moves. (True nondeterminism is essentially unknown in economic game theory; for a use in game theory in computer scientist see \citep[chapter 9]{lavalle06}.) The idea is to define a new `sum of selection functions' operator
\[ \oplus : \J^M_R X \times \J^M_R Y \to \J^M_R (X \times Y) \]
which is analogous to the product of selection functions, but for simultaneous games. Games with nondeterministic strategies are noticeably better behaved than either pure or mixed strategies, with solution spaces carrying more structure.

\item For infinite games, and games with mixed strategies, things are more interesting and difficult. The author is bringing together many ideas from functional programming, topology and category theory to attack these problems. One starting point is that probability distributions form a monad \citep{giry82, erwig06} which carries additional topological structure.
\end{itemize}

\paragraph{Acknowledgments:}
The author is grateful to the anonymous reviewers for their comments, and also gratefully acknowledges EPSRC grant EP/K50290X/1 which funded this work.

\bibliographystyle{plain}
\bibliography{refs}

\end{document}